# Low temperature dynamic freezing and the fragility of ordering in $Tb_2Sn_2O_7$

M. L. Dahlberg [a], M. J. Matthews [a], P. Jiramongkolchai [b], R. J. Cava [b], P. Schiffer [a]

[a] *Department of Physics and Materials Research Institute, Pennsylvania State University, University Park, PA 16803 USA*

[b] *Department of Chemistry, Princeton University, Princeton, NJ 08544 USA*

**Abstract**

We have probed the low temperature magnetic behavior of the ordered spin ice material $Tb_2Sn_2O_7$ through ac magnetic susceptibility measurements of both the pure material and samples with small percentages of Ti substituted on the Sn sublattice. We observe a clear signature for the previously reported ordering transition at $T_C = 850$ mK, as well as evidence for dynamic freezing at temperatures well below $T_C$, confirming the persistence of significant magnetic fluctuations deep in the spin-ordered regime. The long range ordering transition was completely suppressed with as little as 5% Ti for Sn substitution, and 10% Ti substitution resulted in a spin-glass-like spin freezing transition near 250 mK. The results demonstrate that the long range magnetic ordering is surprisingly fragile in this system.

The rare-earth pyrochlore oxides have provided important examples of exotic behavior due to the frustration of magnetic interactions. These materials have the chemical formula $R_2M_2O_7$, with R and M being rare earth and non-magnetic metal ions respectively, each of which is situated on a lattice of corner-sharing tetrahedra. The low temperature magnetic states of these materials include exotic long range ordering, dynamically frozen disordered states that can be either glassy or ice-like, and cooperative paramagnets with strong fluctuations in the low temperature limit [1]. While the low temperature behavior of some frustrated rare earth pyrochlores, such as the spin ices, shows little dependence on the non-magnetic M ion (e.g., $Ho_2Ti_2O_7$ and $Ho_2Sn_2O_7$ [2,3,4,5] or $Dy_2Ti_2O_7$ and $Dy_2Sn_2O_7$ [5,6,7]), $Tb_2Ti_2O_7$ and $Tb_2Sn_2O_7$ have dramatically different low temperature states. The ground state of $Tb_2Ti_2O_7$ is an apparent cooperative paramagnet, in which the spins remain fluctuating to the lowest measured temperatures of 50 mK despite spin-spin interactions with an energy scale set by a Curie-Weiss temperature of -19 K [8] (although there have been recent theoretical predictions of spin-ice-like correlations [9,10]). Low temperature studies of $Tb_2Sn_2O_7$, by contrast, have shown a transition to a long-range ordered state at $T_C$ ~ 850 mK [11]. The ordered state results from effective ferromagnetic interactions between the spins in combination with <111> single-ion spin anisotropy due to the local crystal fields, yielding an apparent ordered spin ice state, i.e., two spins pointing in and two pointing out of each tetrahedron [11,12]. Unlike the canonical spin ice materials, such as $Dy_2Ti_2O_7$ and $Ho_2Ti_2O_7$, the spins in $Tb_2Sn_2O_7$ are canted by ~13.3 degrees off the local <111> axis [13]. Despite the clear indications of spin-ordering, several studies have also reported the presence of significant spin fluctuations well below $T_C$ [14,15,16,17,18].

The primary contrast between the Dy and Ho spin ice systems and the Tb pyrochlores originates in the crystal field level spacing for the rare earth ions. Instead of the 300-350 K gap [19] between the ground state doublet and the first excited state in the Dy and Ho systems, $Tb_2Sn_2O_7$ and $Tb_2Ti_2O_7$ each have been suggested to have a ground state doublet and a second doublet at ~20 K above the ground state [8,13,20]. Recent neutron scattering [21] and heat capacity [22] studies have suggested that this doublet-doublet picture of the lowest crystal field levels, might be altered by a tetragonal distortion below ~20 K that could split the ground state doublets into two singlets.

Here we report measurements of the low temperature ac magnetic susceptibility of pure $Tb_2Sn_2O_7$, as well as samples with disorder introduced by partial substitution of Ti on the Sn site, i.e., of the form $Tb_2Sn_{2-x}Ti_xO_7$. Our data probe the spin system on a longer time scale than previous studies, and demonstrate the existence of low frequency dynamic behavior well below ordering temperature. In the Ti-substituted samples, the introduction of as little as 5% Ti ($x = 0.1$) appears to completely suppress the long range magnetic order. This fragility of the long-range order is rather surprising, since the frustration-induced collective spin states in other rare earth pyrochlores are robust against much higher levels of chemical disorder.

The samples were prepared with standard solid state synthesis techniques, and x-ray diffraction measurements showed the lattice constant to vary linearly with Ti substitution, as expected from Vegard's Law [23]. We measured the dc susceptibility of our samples with a Quantum Design MPMS SQUID magnetometer. We also measured the ac magnetic susceptibility above $T = 1.8$ K using a Quantum Design PPMS with an ac magnetic susceptibility (ACMS) option. At lower temperatures, we measured ac magnetic susceptibility with a custom-built mutual inductance coil susceptometer immersed in helium and thermally anchored to the mixing chamber of a dilution refrigerator. The small oscillating field ($H_{osc} < 1$ Oe) had a variable frequency between $f$ = 10 Hz and 1 kHz.

In figure 1, we plot the high temperature magnetic susceptibility of $Tb_2Sn_2O_7$, measured both as the ac susceptibility and the field derivative of the dc magnetization. As seen in the figure, the data are qualitatively quite similar to those previously published on $Tb_2Ti_2O_7$ [24], and neither material shows any indication of magnetic ordering at temperatures above 1.8 K. The similarities extend to the presence of a slow spin relaxation phenomenon in the presence of a large external field, an effect that was demonstrated previously by our group to be a common feature of similar rare earth magnets [24]. Curie-Weiss fits to higher temperature magnetization data for $Tb_2Sn_{2-x}Ti_xO_7$, x = 0, 0.1 and 0.2, give a Curie-Weiss temperatures of -12.0 ±1.4 K for fits between 50 and 300 K, consistent with previous measurements on $Tb_2Sn_2O_7$ [25].

Figure 2 shows the low temperature ($T < 1.5$ K) ac susceptibility of $Tb_2Sn_2O_7$ in the absence of an applied static magnetic field. We observe a clear feature associated with the transition to long range magnetic order in both the real and imaginary parts of the ac susceptibility, $\chi'(T)$ and $\chi''(T)$. As expected for a long range ordering transition, there is very little frequency dependence to this feature. We do observe an increase in $\chi''(T)$ around 1.2 K, which corresponds in temperature with an increase in ferromagnetic spin-spin correlations [11]. Data taken in the presence of an external static magnetic field up to $H = 0.5$ T are shown in figure 3. Such a field suppresses the magnitude of the features in both $\chi'(T)$ and $\chi''(T)$ and shifts them to higher temperature as expected for ferromagnetic ordering (the inset shows the field dependence of the feature in $\chi'(T)$).

While the long range ordering transition is the most striking feature in our susceptibility data, the temperature dependence of the susceptibility data at temperatures well below the transition is particularly interesting. Careful examination of both $\chi'(T)$ and $\chi''(T)$ reveals complex behavior that could not be discerned from previous measurements taken at a smaller number of temperatures. These features consist of a shoulder in $\chi'(T)$ at T ~ 300 mK, and two distinct peaks in $\chi''(T)$ near 150 mK and 300 mK. We also observed these low temperature features in an independently-prepared sample of $Tb_2Sn_2O_7$, suggesting that they represent generic properties of the material system. While the lowest temperature peak in $\chi''(T)$ appears to have minimal frequency dependence, the higher temperature feature in $\chi''(T)$, corresponding to the broad feature in $\chi'(T)$, has strong frequency dependence in the frequency range of our data. An Arrhenius fit to the peak, shown as the inset to figure 2, gives an activation energy of 1.3 K and characteristic frequency of approximately 83 kHz. The peak shift per decade frequency, $p = (\Delta T_f/T_f\Delta[\log(\omega)])$ [26], calculation gives a value of 0.34. This value excludes a typical spin glass transition, for which one expects $p < 0.1$ [26].

Our data suggest that the spin fluctuations observed in other measurements extend to much longer timescales than probed previously. While our data set does not allow microscopic understanding of these features, the energy scales of the recently suggested crystal field level scheme of 90 and 300 mK [22] do correlate well with the temperatures at which we observe the features. An alternative explanation is that our observed low temperature features are associated with the dynamics of domain walls in the ordered state, although the measured time scales are slower than expected [27].

In figure 4, we compare the low temperature ac susceptibility of the pure $Tb_2Sn_2O_7$ material with that of the $x = 0.1$ and $x = 0.2$ (5% and 10%) Ti-substituted samples. Both samples show a slight increase in $\chi''(T)$ around 900 mK, perhaps associated with some short range ordering, but the long range ordering peak in $\chi'(T)$ is absent from both of the substituted samples. Figure 5 shows the frequency dependence of $\chi'(T)$ and $\chi''(T)$ for the substituted samples. The $x = 0.1$ sample shows only a broad feature near 600 mK in $\chi'(T)$ with weak frequency dependence. The peak in the susceptibility data for the $x = 0.2$ sample, by contrast, exhibits clear features, such as the frequency dependence and the field dependence (data not shown), that are strongly reminiscent of a spin glass transition. In fact, the frequency shift per decade for the $x = 0.1$ sample was p = 0.094, within the range expected for a spin glass [26]. The lack of a similar glassy peak in $\chi'(T)$ of the 5% sample suggests that a minimum level of substitution-induced disorder is necessary for this glassy state to be fully realized.

The suppression of the ordered spin ice transition with minimal Ti-substitution is surprising, since the spin ice state in Ho and Dy materials has been shown to be very robust against the introduction of structural disorder. The introduction of a significant amount of antimony on the Sn sublattice of $Dy_2Sn_2O_7$ [6], for example, created structural and charge disorder, but the spin ice state remained effectively unchanged. The introduction of non-magnetic ions on the magnetic sites in other spin ice systems has yielded similar results, in that the spin ice physics remained essentially intact [28,29,30]. Even magnetic dilution studies of $Tb_2Ti_2O_7$, where the $Tb^{3+}$ ions were replaced with non-magnetic $Y^{3+}$, showed little change in the magnetic properties down to T = 2 K [31]. A comparison to another model rare earth Ising system $LiHo_xY_{1-x}F_4$, also shows this behavior to be anomalous, since that system shows ferromagnetism over a wide range of disorder [32].

One possible cause for the disruption of long range order by Ti substitution is simply the change in lattice size that the substitution induces. The difference in size between the Sn and Ti ions implies that the substitution will alter both the dipolar and exchange interactions of nearby Tb ions, and thus one might expect consequences for the low temperature cooperative spin state. Studies of $Tb_2Sn_2O_7$ under applied isotropic pressure and uniaxial stress showed a melting of the ordered state, and the emergence of both spin liquid behavior and a new k = 001 antiferromagnetic ordering similar to that seen in $Tb_2Ti_2O_7$ under pressure [33]. As such, a melting of the $Tb_2Sn_2O_7$ spin ice ordering by chemically changing the average lattice spacing is certainly plausible. Alternatively, the suppression of ordering through the substitutions could be due to local alterations of the crystal field levels of the Tb ions by substitution of Ti for Sn. In the usual doublet-doublet picture of the low lying crystal field levels, it has also been shown that the two low-lying doublets of $Tb_2Ti_2O_7$ and $Tb_2Sn_2O_7$ are inverted with respect to each other, even though the energy spacing is very similar [13]. Presumably the local disorder of the Ti substitution is sufficient to disrupt the crystal field levels of all Tb ions neighboring an introduced Ti, and therefore would have a significant effect on the ordering of the moments. This disorder, combined with disorder in the interactions between moments, is also likely to be the root cause of the spin-glass-like low temperature behavior seen in the $x = 0.2$ sample.

Regardless of the physical origin of the suppression of the ordering through Ti substitution for Sn, the demonstrated fragility of the ordered state indicates that it results from a precise balancing of different interactions in this material system. Further studies, such as EXAFS or inelastic neutron scattering could be performed to determine the nature of the structural disorder and the impact of the substitution-induced disorder on the low lying crystal field levels of the Ti-substituted samples. Similarly, a more detailed study of the effects of pressure could discern whether disorder or simply the lattice change is responsible for the suppression of ordering. Regardless of the ultimate basis for the fragility of the ordered spin ice state, the dynamic behavior we observe in both pure and Ti-substituted $Tb_2Sn_2O_7$ indicates that this model system provides a new paradigm for frustrated magnetism, being on the cusp between an ordered and a fluctuating phase.

We gratefully acknowledge useful discussion with F. Bert, P. Mendels and J. S. Gardner. This research was supported in part by the NSF Grant No. DMR-070158.

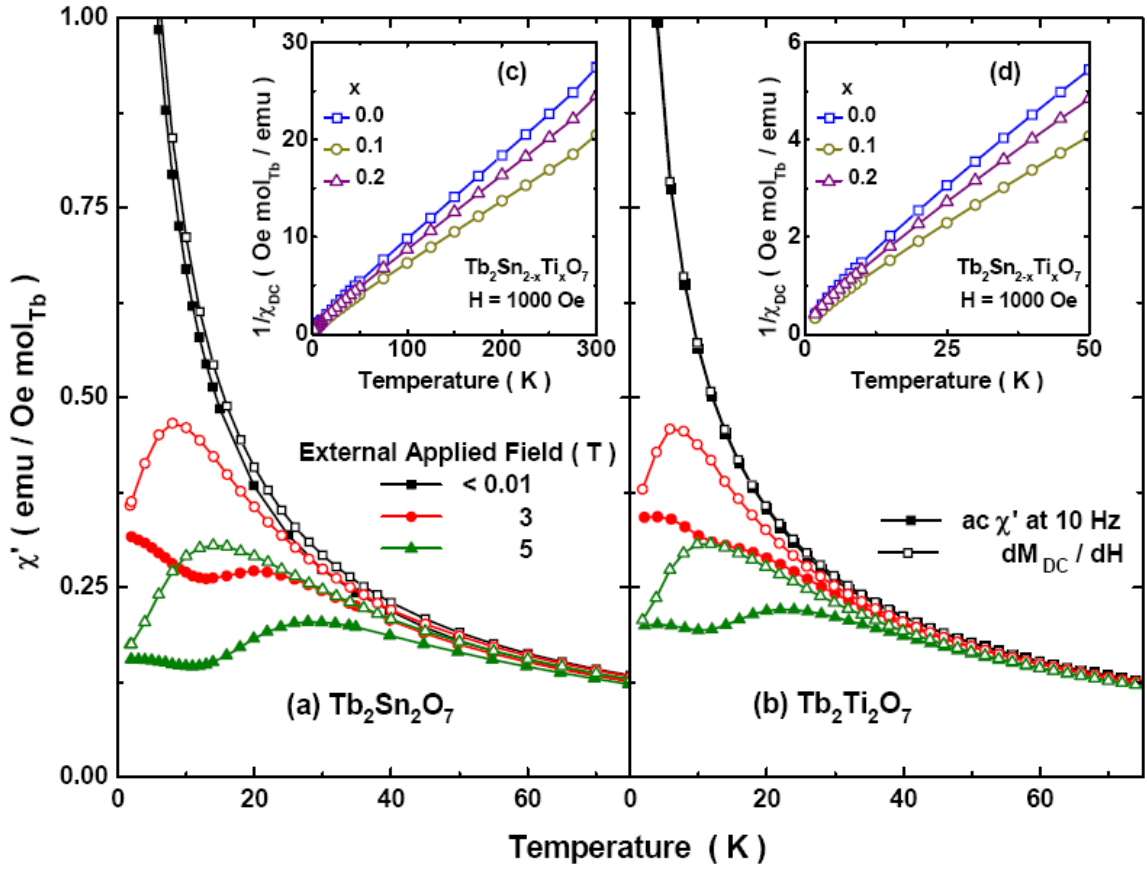

Figure 1. (Color online) High temperature ac susceptibility (closed symbols) and dM/dH$_{DC}$ (open symbols) data on (a) Tb$_2$Sn$_2$O$_7$ and (b) Tb$_2$Ti$_2$O$_7$ (reproduced from [24]). The similarities are indicative of similarities in the CF level spacing of the two systems. The insets show the inverse dc susceptibility of the three different samples studied in the present work; (c) over the full temperature range from 1.8 – 300 K and (d) over the lover temperature range from 1.8 – 50 K.

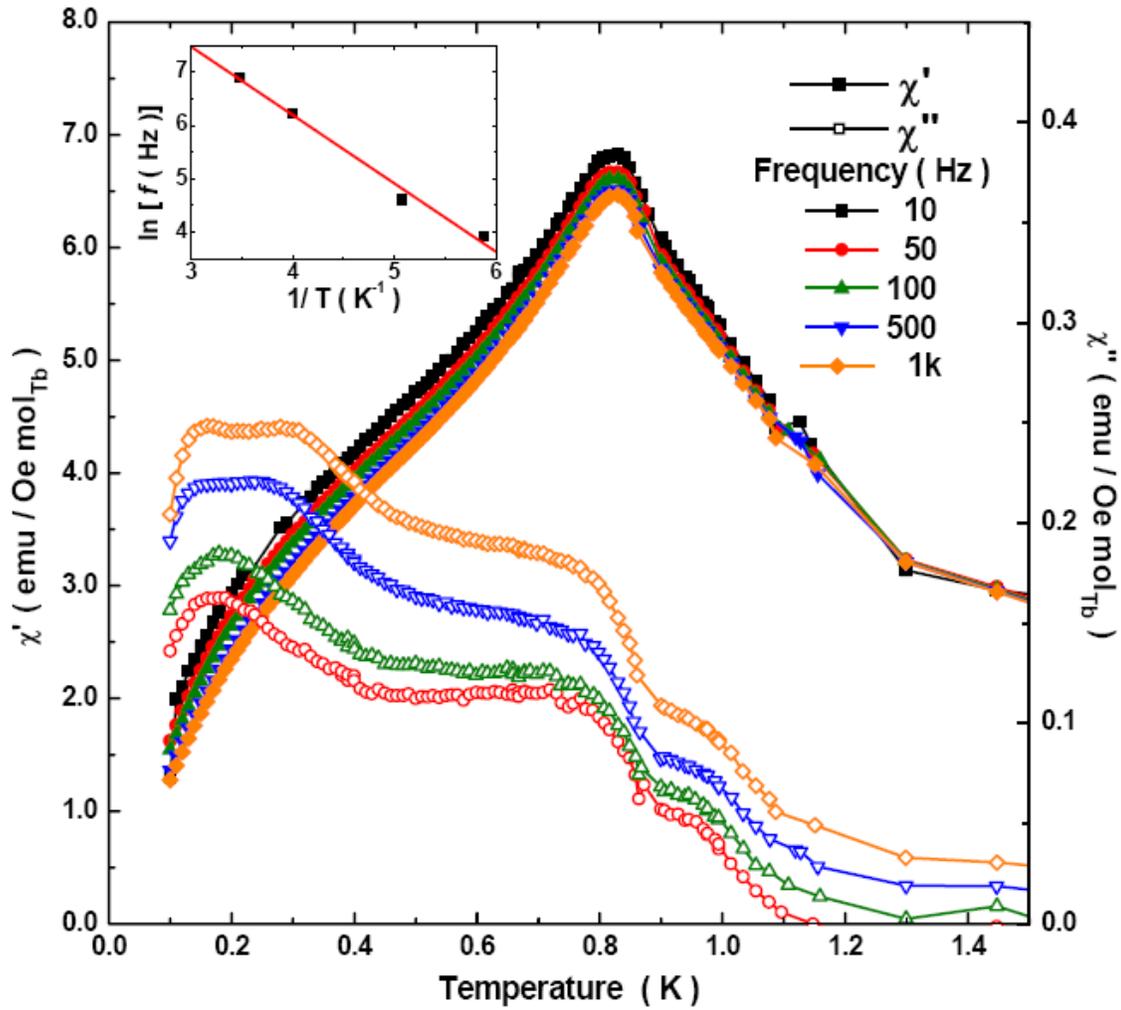

Figure 2. (Color online) The low temperature ac susceptibility of $Tb_2Sn_2O_7$ in zero external applied field. Closed symbols show $\chi'(T)$; open symbols show $\chi''(T)$. The inset shows an Arrhenius fit of the frequency dependence of the 300 mK feature.

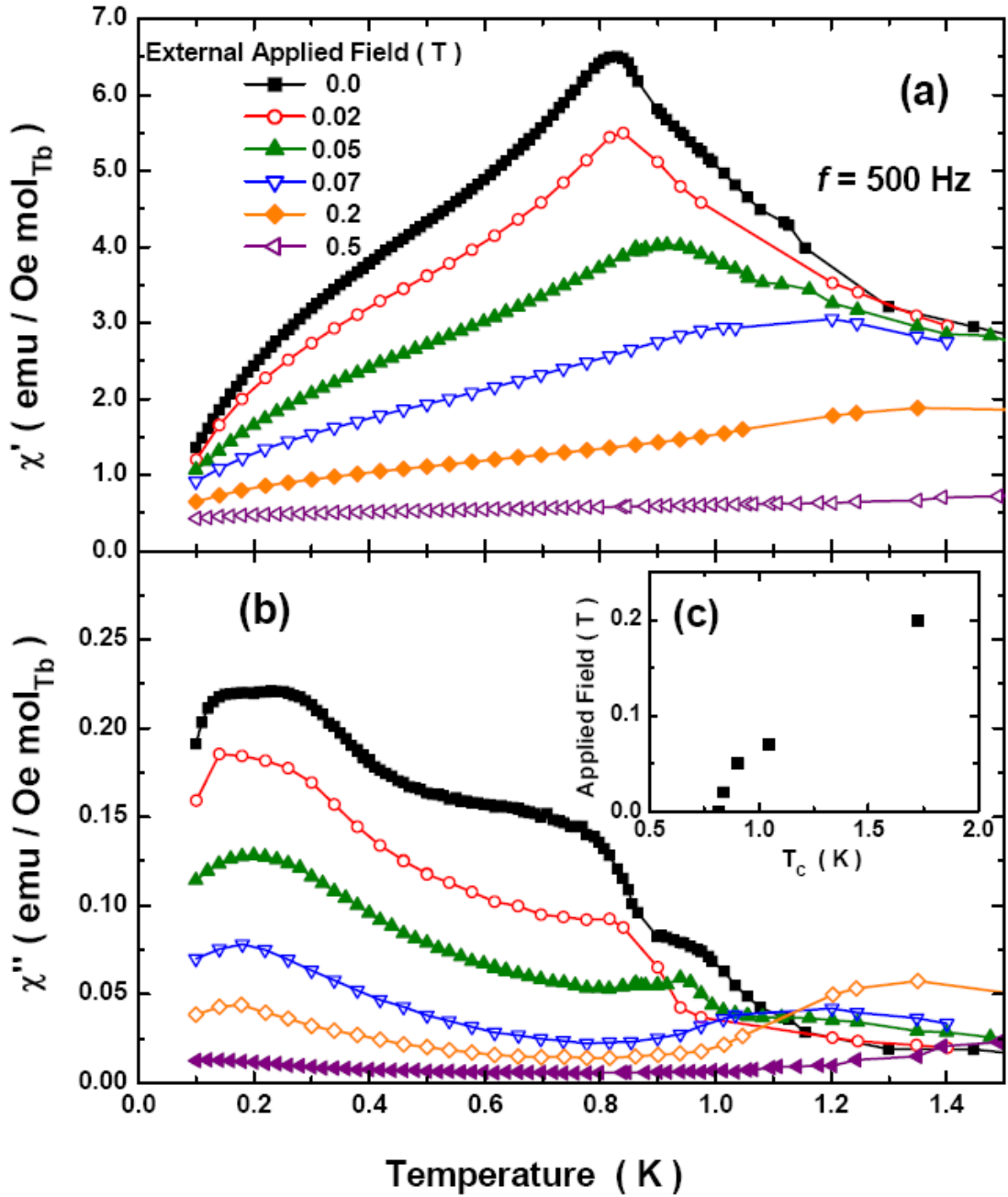

Figure 3. (Color online) Low temperature data taken on $Tb_2Sn_2O_7$ in various external applied fields (a) $\chi'(T)$ clearly shows the ordering feature moving to higher temperature with increasing field as expected. (b) $\chi''(T)$ shows the field dependence of the two low temperature features. The inset (c) shows the dependence of the peak position on the external applied field.

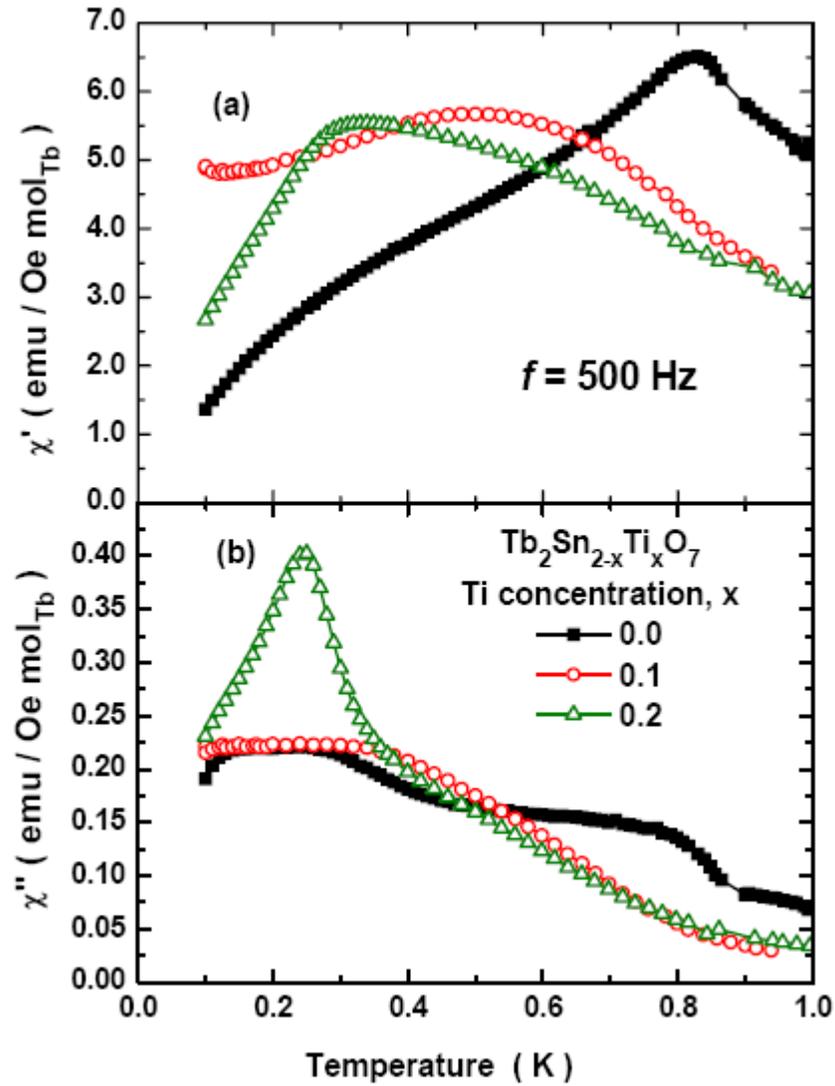

Figure 4. (Color online) Dilution refrigeration ac magnetic susceptibility measurements on the three samples studied: the pure $Tb_2Sn_2O_7$ and the two Ti-substituted samples in zero external applied field.

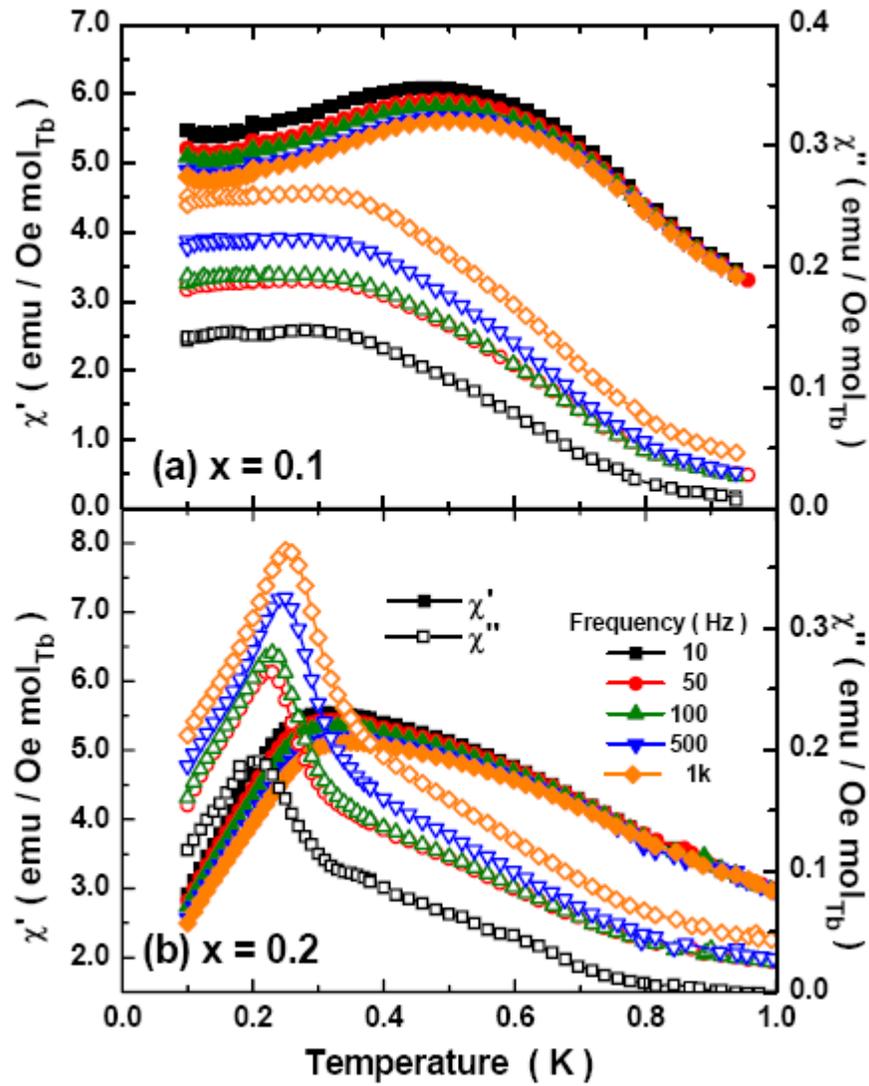

Figure 5. (Color online) Frequency dependence of the $x = 0.1$ and $x = 0.2$ samples showing the suppression of ordering and the emergence of spin-glass-like behavior for $x = 0.2$. Closed symbols show $\chi'(T)$; open symbols show $\chi''(T)$.